\documentstyle[11pt,psfig,twoside]{article}

\begin{document}

\newcommand{\dd}{\,{\rm d}}
\newcommand{\ie}{{\it i.e.},\,}
\newcommand{\etal}{{\it et al.\ }}
\newcommand{\eg}{{\it e.g.},\,}
\newcommand{\cf}{{\it cf.\ }}
\newcommand{\vs}{{\it vs.\ }}
\newcommand{\zdot}{\makebox[0pt][l]{.}}
\newcommand{\up}[1]{\ifmmode^{\rm #1}\else$^{\rm #1}$\fi}
\newcommand{\dn}[1]{\ifmmode_{\rm #1}\else$_{\rm #1}$\fi}
\newcommand{\upd}{\up{d}}
\newcommand{\uph}{\up{h}}
\newcommand{\upm}{\up{m}}
\newcommand{\ups}{\up{s}}
\newcommand{\arcd}{\ifmmode^{\circ}\else$^{\circ}$\fi}
\newcommand{\arcm}{\ifmmode{'}\else$'$\fi}
\newcommand{\arcs}{\ifmmode{''}\else$''$\fi}
\newcommand{\MS}{{\rm M}\ifmmode_{\odot}\else$_{\odot}$\fi}
\newcommand{\RS}{{\rm R}\ifmmode_{\odot}\else$_{\odot}$\fi}
\newcommand{\LS}{{\rm L}\ifmmode_{\odot}\else$_{\odot}$\fi}

\newcommand{\Abstract}[2]{{\footnotesize\begin{center}ABSTRACT\end{center}
\vspace{1mm}\par#1\par
\noindent
{~}{\it #2}}}

\newcommand{\TabCap}[2]{\begin{center}\parbox[t]{#1}{\begin{center}
  \small {\spaceskip 2pt plus 1pt minus 1pt T a b l e}
  \refstepcounter{table}\thetable \\[2mm]
  \footnotesize #2 \end{center}}\end{center}}

\newcommand{\TableSep}[2]{\begin{table}[p]\vspace{#1}
\TabCap{#2}\end{table}}

\newcommand{\FigCap}[1]{\footnotesize\par\noindent Fig.\  %
  \refstepcounter{figure}\thefigure. #1\par}

\newcommand{\TableFont}{\footnotesize}
\newcommand{\TableFontIt}{\ttit}
\newcommand{\SetTableFont}[1]{\renewcommand{\TableFont}{#1}}

\newcommand{\MakeTable}[4]{\begin{table}[htb]\TabCap{#2}{#3}
  \begin{center} \TableFont \begin{tabular}{#1} #4 
  \end{tabular}\end{center}\end{table}}

\newcommand{\MakeTableSep}[4]{\begin{table}[p]\TabCap{#2}{#3}
  \begin{center} \TableFont \begin{tabular}{#1} #4 
  \end{tabular}\end{center}\end{table}}

\newenvironment{references}%
{
\footnotesize \frenchspacing
\renewcommand{\thesection}{}
\renewcommand{\in}{{\rm in }}
\renewcommand{\AA}{Astron.\ Astrophys.}
\newcommand{\AAS}{Astron.~Astrophys.~Suppl.~Ser.}
\newcommand{\ApJ}{Astrophys.\ J.}
\newcommand{\ApJS}{Astrophys.\ J.~Suppl.~Ser.}
\newcommand{\ApJL}{Astrophys.\ J.~Letters}
\newcommand{\AJ}{Astron.\ J.}
\newcommand{\IBVS}{IBVS}
\newcommand{\PASP}{P.A.S.P.}
\newcommand{\Acta}{Acta Astron.}
\newcommand{\MNRAS}{MNRAS}
\renewcommand{\and}{{\rm and }}
\section{{\rm REFERENCES}}
\sloppy \hyphenpenalty10000
\begin{list}{}{\leftmargin1cm\listparindent-1cm
\itemindent\listparindent\parsep0pt\itemsep0pt}}%
{\end{list}\vspace{2mm}}

\def\TYLDA{~}
\newlength{\DW}
\settowidth{\DW}{0}
\newcommand{\dw}{\hspace{\DW}}

\newcommand{\refitem}[5]{\item[]{#1} #2%
\def\REFARG{#3}\ifx\REFARG\TYLDA\else, {\it#3}\fi
\def\REFARG{#4}\ifx\REFARG\TYLDA\else, {\bf#4}\fi
\def\REFARG{#5}\ifx\REFARG\TYLDA\else, {#5}\fi.}

\newcommand{\Section}[1]{\section{#1}}
\newcommand{\Subsection}[1]{\subsection{#1}}
\newcommand{\Acknow}[1]{\par\vspace{5mm}{\bf Acknowledgements.} #1}
\pagestyle{myheadings}

\def\thefootnote{\fnsymbol{footnote}}
\begin{center}
{\Large\bf The Optical Gravitational Lensing Experiment.\\
\vskip3pt
Eclipsing Binary Stars in the Small Magellanic Cloud
\footnote{Based on  observations obtained with the 1.3~m Warsaw
telescope at the Las Campanas  Observatory of the Carnegie Institution
of Washington.}}
\vskip1cm
{\bf
A.~~U~d~a~l~s~k~i$^1$,~~I.~~S~o~s~z~y~{\'n}~s~k~i$^1$,
~~M.~~S~z~y~m~a~{\'n}~s~k~i$^1$,~~M.~~K~u~b~i~a~k$^1$,
~~G.~~P~i~e~t~r~z~y~\'n~s~k~i$^1$,
~~P.~~W~o~\'z~n~i~a~k$^2$,~~ and~~K.~~\.Z~e~b~r~u~\'n$^1$}
\vskip3mm
{$^1$Warsaw University Observatory, Al.~Ujazdowskie~4, 00-478~Warszawa, Poland\\
e-mail: (udalski,soszynsk,msz,mk,pietrzyn,zebrun)@sirius.astrouw.edu.pl\\
$^2$ Princeton University Observatory, Princeton, NJ 08544-1001, USA\\
e-mail: wozniak@astro.princeton.edu}
\vskip1cm
\end{center}

\Abstract{We present the catalog of 1459 eclipsing binary stars detected
in the  central 2.4 square degree area of the Small Magellanic Cloud
during the  OGLE-II microlensing search. The sample includes objects
brighter than  ${I\approx 20}$~mag with periods ranging from about 0.3
to 250 days. The  average completeness of the catalog is about 80\%.
Statistics of the  entire sample and well detached systems, suitable for
distance determination, are also presented.

The catalog, finding charts and {\it BVI} photometry data for all
objects  are  available from the OGLE Internet archive.}

\Section{Introduction}
Eclipsing binary stars are among the oldest variable stars recognized
and  correctly interpreted by astronomers. In 1780s  Pigott and
Goodricke observed  $\beta$~Per (Algol) and explained its variability by
binarity of the system  in which two components obscure periodically
each other. For years methods of  analysis of eclipsing star light
curves were developed and improved. The  binary systems became one of
the most important sources of information about  stellar parameters like
radii, masses, luminosities etc. Also theory of evolution of binary 
stars, in particular close systems exchanging mass during evolution, was
 developed during the past decades. Thus, eclipsing systems seem to be a
group  which is well understood, although somewhat off  the main
astrophysical  interest. 

However, eclipsing binary stars might once again become one of the most 
important objects in the near Universe. They seem to be the most
promising  candidates for standard candles  allowing determination of
distances within  the Local Group with so far unprecedented accuracy of
a few percent  (Paczy{\'n}ski 1997). 

The principle of the method was first recognized by Stebbins (1910).
With  accurate photometry and spectroscopy absolute dimensions of both
components  can precisely be derived and, together with accurately
determined  temperatures, \ie surface brightness of each component,
emitted fluxes can be  calculated. This should give precise distance
when compared with the flux  observed at the Earth. The method is
essentially geometric, free from possible  population effects
influencing other standard candles. The parameters of   components
should be determined with high precision, therefore only well  detached
systems should be used. A few percent accuracy of distance 
determination should be achieved when good quality observations are
available. 

Eclipsing binaries are very common but difficult to detect because very
good  sampling of the light curve is necessary to detect eclipses.
Fortunately large  samples of these objects emerge as by-products of
microlensing searches, \eg  in the Large Magellanic Cloud (Grison \etal
1994, Alcock \etal 1997) or  Galactic bulge (Udalski \etal 1997a).
Several eclipsing binaries in M31 and  M33 have recently been discovered
by DIRECT project searching for eclipsing  systems in these galaxies
(Kaluzny \etal 1998, Stanek \etal 1998). 

First successful application of the eclipsing binary for distance 
determination to the LMC was already presented with HV2274 system
(Guinan  \etal 1998, Udalski \etal 1998b). Results of these studies seem
to confirm  shorter distance scale to the LMC, ${(m-M)_{\rm
LMC}\approx18.2-18.3}$~mag,  than generally accepted value of
${(m-M)_{\rm LMC}\approx18.5}$~mag making an  important step toward
ending the long lasting dispute on the LMC distance. The  distance to
the LMC is crucial for modern astrophysics, as the extragalactic 
distance scale is tied to the LMC distance. We believe that eclipsing
binaries  will provide accurate absolute distances to many objects
within the Local  Group enabling precise calibration of other standard
candles. 

This paper is aimed at providing a list and photometry of eclipsing
binaries  detected during the OGLE-II microlensing search (Udalski,
Kubiak and  Szyma{\'n}ski 1997b) in the Small Magellanic Cloud. The
catalog of eclipsing  binaries contains 1459 objects with full variety
of eclipsing systems. The  sample is reasonably complete, allowing
statistical analysis. It should  provide a good material for testing
theory of evolution of binary systems as  well as studying the SMC
evolution, star formation etc. 

Many of the eclipsing systems presented in this paper are well detached 
systems well suited for distance determination. With the largest 6--8~m
class  telescopes being currently constructed in the southern
hemisphere, it should  be possible to obtain high resolution spectra of
all these, even quite faint,  objects. Eclipsing binaries should then
provide not only precise mean distance  to the SMC but, also, should
allow to study geometry and  spatial distribution  of stars in the SMC. 

\vspace*{-4pt}
\Section{Observations}
\vspace*{-2pt}
All observations presented in this paper were carried out with the 1.3-m
 Warsaw telescope at the Las Campanas Observatory, Chile, which is
operated by  the Carnegie Institution of Washington, during the second
phase of the OGLE  experiment. The telescope was equipped with the
"first generation" camera with  a SITe ${2048\times2048}$ CCD detector.
The pixel size was 24~$\mu$m giving  the 0.417 arcsec/pixel scale.
Observations of the SMC were performed in the  "slow" reading mode of
the CCD detector with the gain 3.8~e$^-$/ADU and  readout noise about
5.4~e$^-$. Details of the instrumentation setup can be  found in
Udalski, Kubiak and Szyma{\'n}ski (1997b). 

Observations of the SMC started on June~26, 1997. As the microlensing
search  is planned to last for a few years, observations of selected
fields will be  continued during following seasons. In this paper we
present data collected up to  November~27, 1998. Observations are
obtained in the standard {\it BVI}-bands  with majority of measurements
made in the {\it I}-band. 

Photometric data collected in the course of the first observing season
of the  SMC for 11 fields (SMC$\_$SC1--SMC$\_$SC11) covered about 2.4
square degree  of the central parts  of the SMC and were used to
construct the {\it BVI} maps of  the SMC (Udalski \etal 1998a). The
reader is referred to that paper for more  details about  methods of
data reduction, tests on quality of photometric data,  astrometry etc. 

\vspace*{-4pt}
\Section{Search for Eclipsing Binary Stars}
\vspace*{-2pt}
Search for variable objects in 11 SMC fields (Table~1) was performed
using  observations in the {\it I}-band in which majority of
observations was  obtained. Typically about 140--180 epochs were
available for each analyzed  object with the lower limit set to~50. The
mean {\it I}-band magnitude of  objects was limited to ${I<20}$~mag.
Candidates for variable stars were  selected based on comparison of the
standard deviation of all individual  measurements of a star with
typical standard deviation for stars of similar  brightness. Light
curves of selected candidates were then searched for  periodicity using
the AoV algorithm (Schwarzenberg-Czerny 1989) which was  successfully
applied in earlier, OGLE-I variable stars searches (\eg Udalski  \etal
1997a). The period search was limited to the range 0.1--100~days. 
However, a few objects were identified with $P/2$ period and their true
period  exceeds that upper limit. 

%\vspace*{-20pt}
\MakeTable{lcc}{12.5cm}{Equatorial coordinates of the SMC fields}
{
\hline
\noalign{\vskip3pt}
\multicolumn{1}{c}{Field} & RA (J2000)  & DEC (J2000)\\
\hline
\noalign{\vskip3pt}
SMC$\_$SC1  & 0\uph37\upm51\ups & $-73\arcd29\arcm40\arcs$\\
SMC$\_$SC2  & 0\uph40\upm53\ups & $-73\arcd17\arcm30\arcs$\\
SMC$\_$SC3  & 0\uph43\upm58\ups & $-73\arcd12\arcm30\arcs$\\
SMC$\_$SC4  & 0\uph46\upm59\ups & $-73\arcd07\arcm30\arcs$\\
SMC$\_$SC5  & 0\uph50\upm01\ups & $-73\arcd08\arcm45\arcs$\\
SMC$\_$SC6  & 0\uph53\upm01\ups & $-72\arcd58\arcm40\arcs$\\
SMC$\_$SC7  & 0\uph56\upm00\ups & $-72\arcd53\arcm35\arcs$\\
SMC$\_$SC8  & 0\uph58\upm58\ups & $-72\arcd39\arcm30\arcs$\\
SMC$\_$SC9  & 1\uph01\upm55\ups & $-72\arcd32\arcm35\arcs$\\
SMC$\_$SC10 & 1\uph04\upm51\ups & $-72\arcd24\arcm45\arcs$\\
SMC$\_$SC11 & 1\uph07\upm45\ups & $-72\arcd39\arcm30\arcs$\\  
\hline}

Light curves of all objects revealing statistically significant periodic
signal were then displayed and visually inspected to avoid spurious 
detections. Eclipsing binaries were extracted from the entire sample
when  their light curve displayed either evident narrow eclipses or
round shape  light curve with wide eclipses of different depth or with
sharp minima. This  should reject ellipsoidal variable stars with
sinusoidal light curves from our  sample of eclipsing systems. Although
the ellipsoidal stars variability is  caused by similar reasons to
eclipsing stars -- orbital motion of components,  they can easily be
confused with variable stars of other types which would  make our sample
inhomogeneous. 

In the next step periods of selected eclipsing star candidates were
tuned to  smooth the eclipse shape which is very sensitive to period
inaccuracies. Then  brightness of each object outside the eclipse was
determined. To derive  colors, observations in the {\it B} and {\it
V}-bands which are less numerous  than in the {\it I}-band  were first
folded with the system period and then  brightness outside eclipse was
determined. Accuracy of colors is about  0.03~mag and 0.02~mag for
${B-V}$ and ${V-I}$, respectively. Accuracy of  period is equal to about
${1\times10^{-4}P}$. 

Each of the analyzed SMC fields overlaps with neighboring fields for 
calibration purposes. Therefore several eclipsing stars located in the 
overlapping regions were detected twice. We decided not to remove them
from  the final list of objects because their measurements are
independent in both  fields. 68 such stars were detected and we provide
cross-reference list to  identify them. 

\Section{Catalog of Eclipsing Binary Stars}
Eclipsing binary stars detected by OGLE-II in the SMC fields  are listed
in Tables~2--12. They contain the identification number in the field
database, equatorial  coordinates (J2000), period in days, zero epoch
corresponding to the deeper  eclipse, {\it I}-band magnitude and ${B-V}$
and ${V-I}$ colors. Photometry  corresponds to the maximum of
brightness. 

All tables contain 1527 stars. 68 of them were identified twice in
overlapping  regions between neighboring fields. Table~13 provides
cross-reference list to  identify these objects.

Appendices A--K present {\it I}-band light curves of all objects
included in  the catalog. The ordinate is phase with 0.0 value
corresponding to the deeper  eclipse. Abscissa is {\it I}-band
magnitude. Light curve is repeated twice for  clarity.

Finding charts are not presented in this paper but they are  available
from the OGLE Internet archive (see below). 

\begin{figure}[htb]
\centerline{\psfig{figure=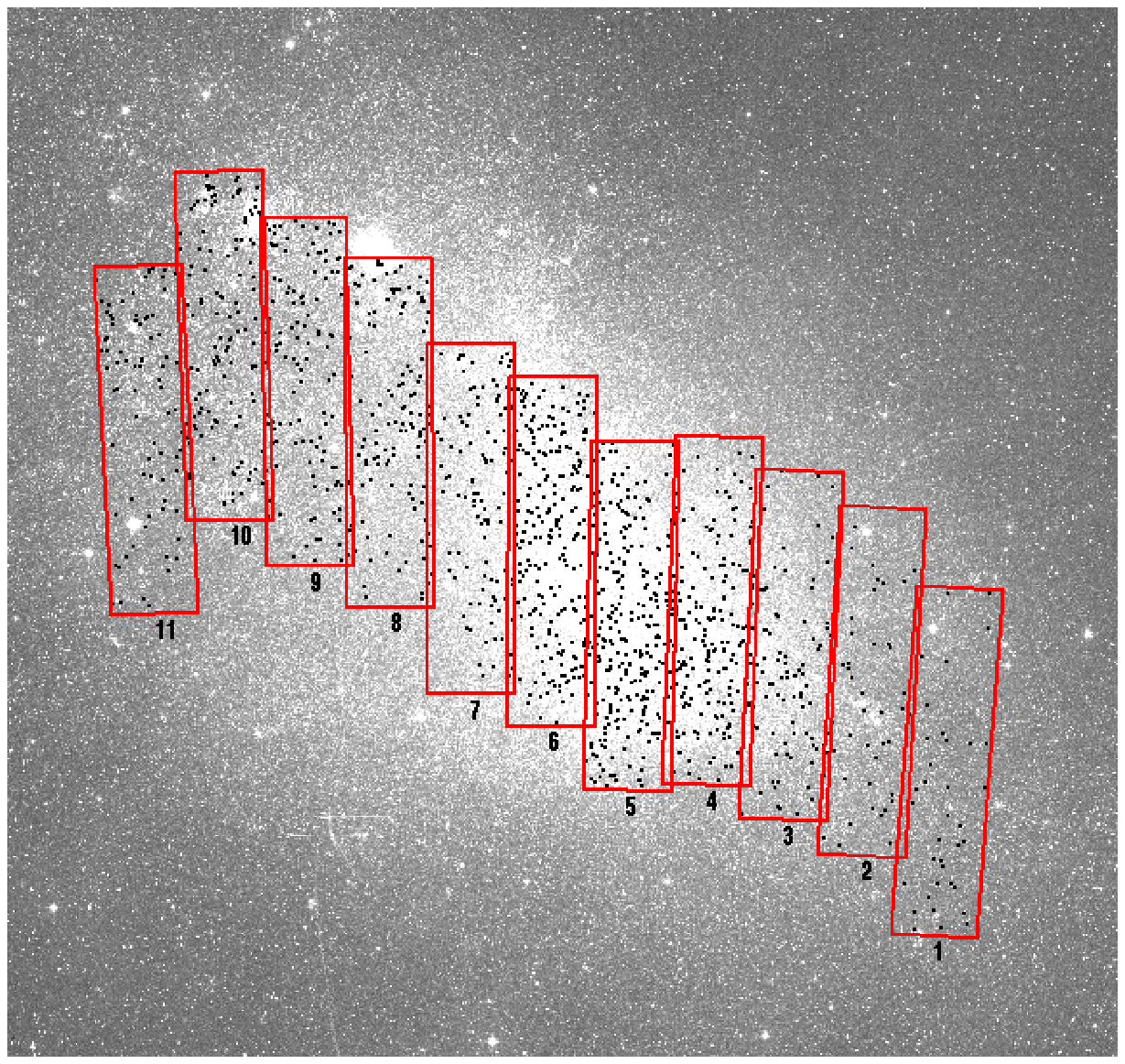,clip=}}
\vskip12pt
\FigCap{OGLE-II fields in the SMC. Black dots indicate positions of
eclipsing stars. North is up and East to the left.}
\end{figure} 

\Section{Discussion}
We present 1459 eclipsing stars located in the central regions of the
SMC.  Although we did not attempt to identify presented stars with
previously known  from photographic observations we suspect that the
vast majority of these  objects are newly discovered eclipsing systems.
The sample contains a wide range of  eclipsing objects -- from well
detached Algol systems, through semi-detached  objects to $\beta$~Lyrae
stars. Many systems reveal eccentric orbits with  possible apsidal
motion. Some of the objects, in particular those with the  longest
periods, might have been misclassified because few cycles were covered 
during the presented period of observations. For example, we included in
the  catalog a strange object, OGLE SMC$\_$SC5 208025, which displays
eclipse-like  features in the light curve superimposed on other
variability. 

Quality of photometry varies from object to object mostly depending on 
brightness of the star and also on  crowding of the field. For many stars 
quality of light curve is sufficient for any application including distance 
determination. For fainter objects more precise data might be 
necessary. However, knowing the ephemeris of such  star one can optimize 
additional observations. 

Fig.~1 presents the picture of the SMC from the Digitized Sky Survey
CD-ROMs  with contours of the OGLE-II fields. Positions of the eclipsing
binaries are  marked with black dots. The stars are distributed more or
less proportionally  to the stellar density with the largest
concentration in the fields  SMC$\_$SC4--SMC$\_$SC6, \ie at the SMC
center. 

\begin{figure}[htb]
\vskip-4mm
\centerline{\psfig{figure=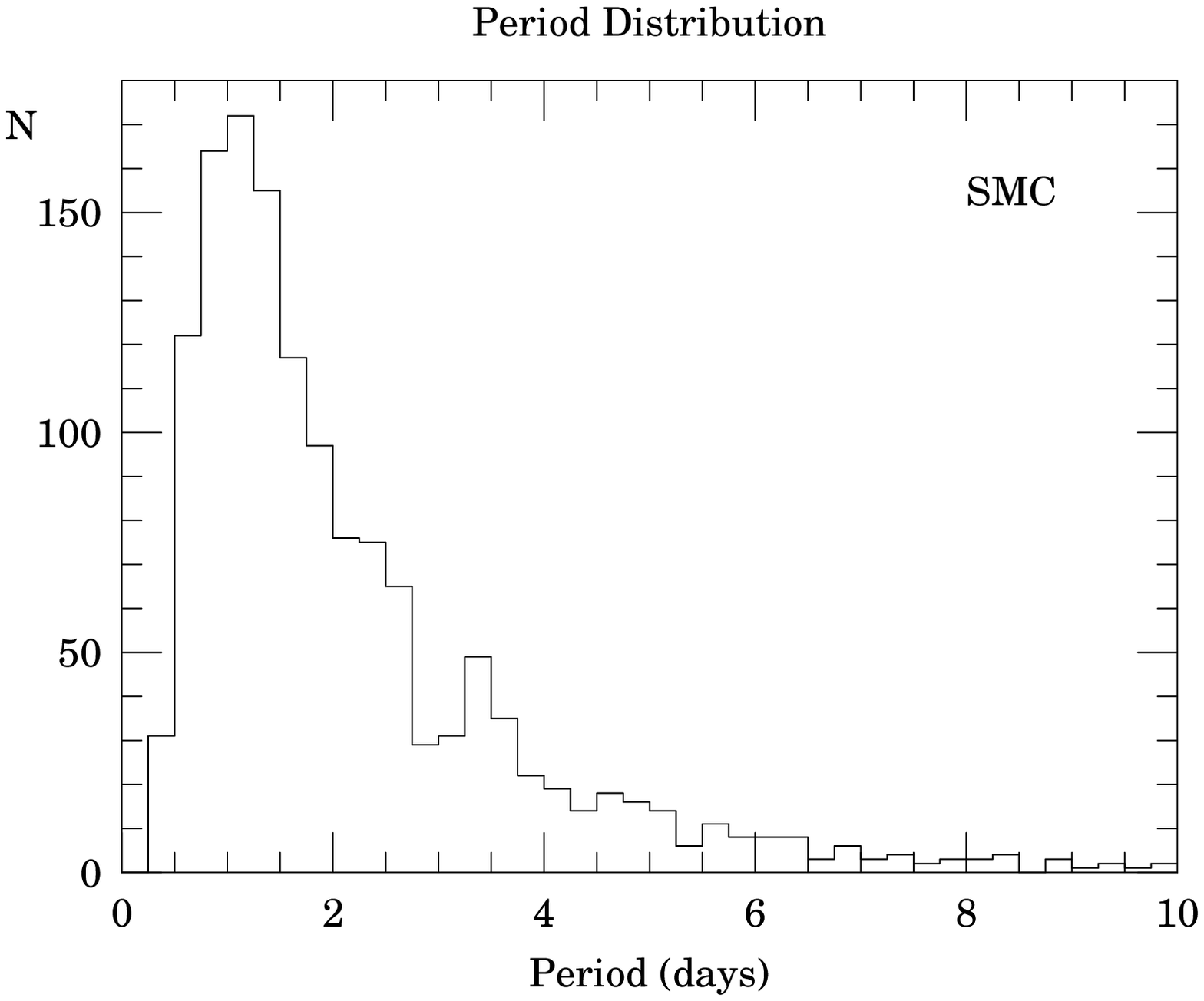,bbllx=0pt,bblly=0pt,bburx=540pt,bbury=440pt,width=10cm,clip=}}
\vskip-8mm
\FigCap{Histogram of orbital periods of eclipsing binaries in 0.25 day
bins. Additional 128 objects have periods longer than 10 days.}
\end{figure} 

\begin{figure}[htb]
\centerline{\psfig{figure=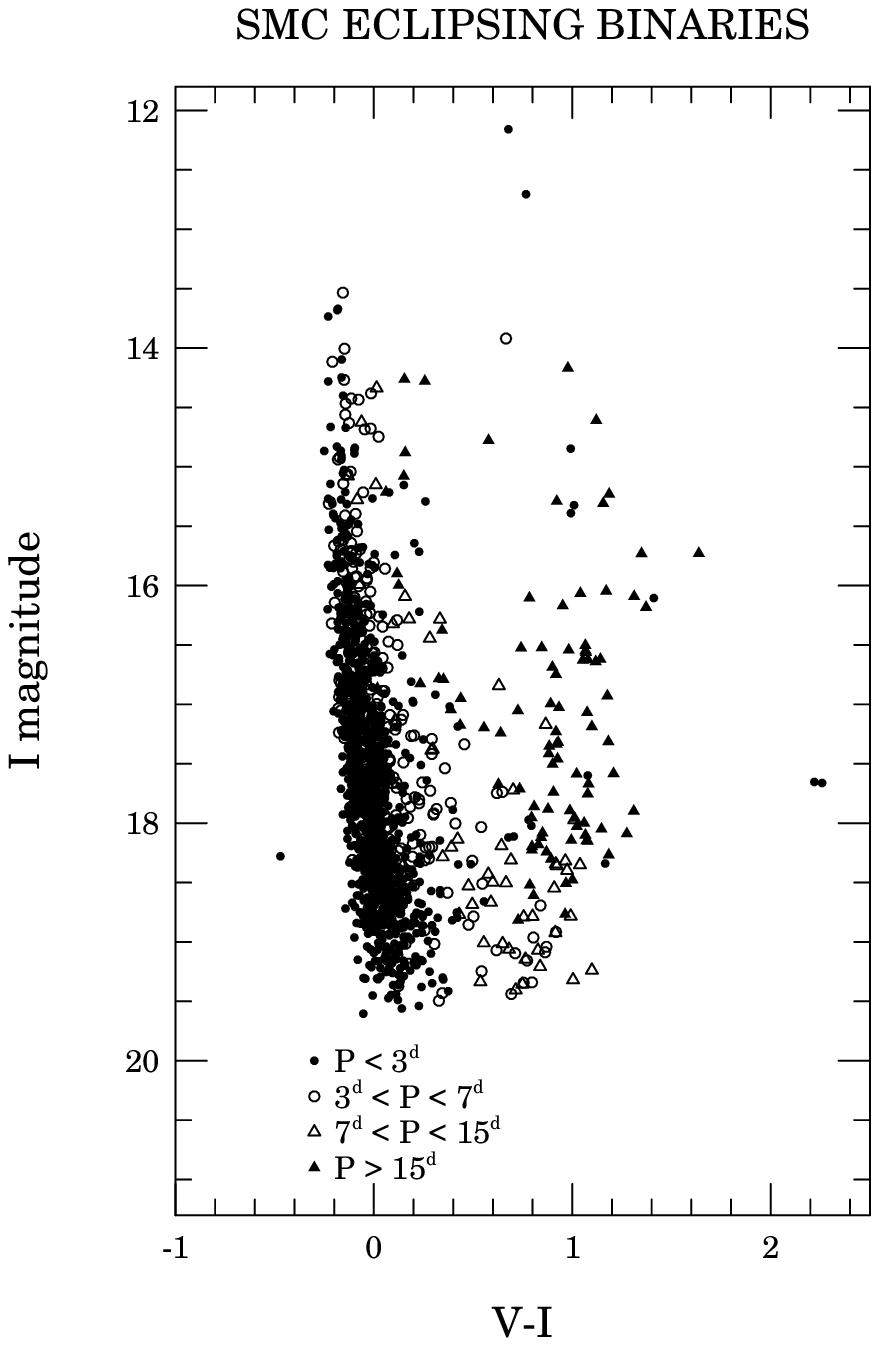,bbllx=120pt,bblly=100pt,bburx=420pt,bbury=510pt,height=12cm,clip=}}
\FigCap{Color-magnitude diagram of eclipsing stars in the SMC. Different
symbols mark positions of stars with short, medium, long and very-long
periods.}
\end{figure} 
Fig.~2 shows the histogram of orbital periods of all stars from the 
catalog in 0.25~day bins from 0.0 to 10 days. Additional 128 objects
with  periods above 10 days are distributed more or less uniformly. As
one can  expect, the majority of stars are short period eclipsing
systems with the  typical orbital period around 1.1~day. The number of
systems falls rapidly  with increasing period.

Fig.~3 presents {\it I} \vs ${V-I}$ color-magnitude diagram for all
stars from  the catalog. Different symbols mark short (${P<3\upd}$),
medium  (${3\upd<P<7\upd}$), long (${7\upd<P<15\upd}$) and very-long
(${P>15\upd}$)  period systems. Fig.~3 indicates that practically all
eclipsing systems from  the catalog belong to the SMC. There are very
few foreground systems. Eclipsing  systems with short and medium periods
belong to the young population located  in the main sequence. Long
period stars are usually lower giant branch stars  or evolved main
sequence stars while very-long period stars populate upper  part of the
red giant branch. The number of giant branch eclipsing systems is  much
smaller than shorter period stars because such wide systems are much
more  rare. Inclination of orbit of those stars must be much closer to
90 deg than for   close binary systems. Also, the efficiency of
detection for very-long period  systems is certainly lower due to
smaller number of cycles covered. 

Figs.~1--3 indicate that the eclipsing stars from the catalog constitute
in  many respects a very representative  sample of stars from the SMC.
The  question is how complete is this sample. In general, the
completeness is a  function of magnitude, quality of photometry, period
etc. We attempted to  determine the mean completeness of our catalog
based on objects detected in  overlapping regions of neighboring fields.
Based on astrometric solutions for  each field we selected eclipsing
stars which should be detected in neighboring  fields and compared these
objects with actually detected stars. 20 such tests  were possible
between our 11 fields. 170 stars should be theoretically paired  up. In
practice 136 pairs were found, yielding the mean completeness of our 
catalog equal to 80\%. It should be noted that this is certainly a lower
limit  as the edge regions of each field are affected by non-perfect
pointing of the  telescope leading to effectively smaller number of
observations there. 

\begin{figure}[htb]
%\vskip-2mm
\centerline{\psfig{figure=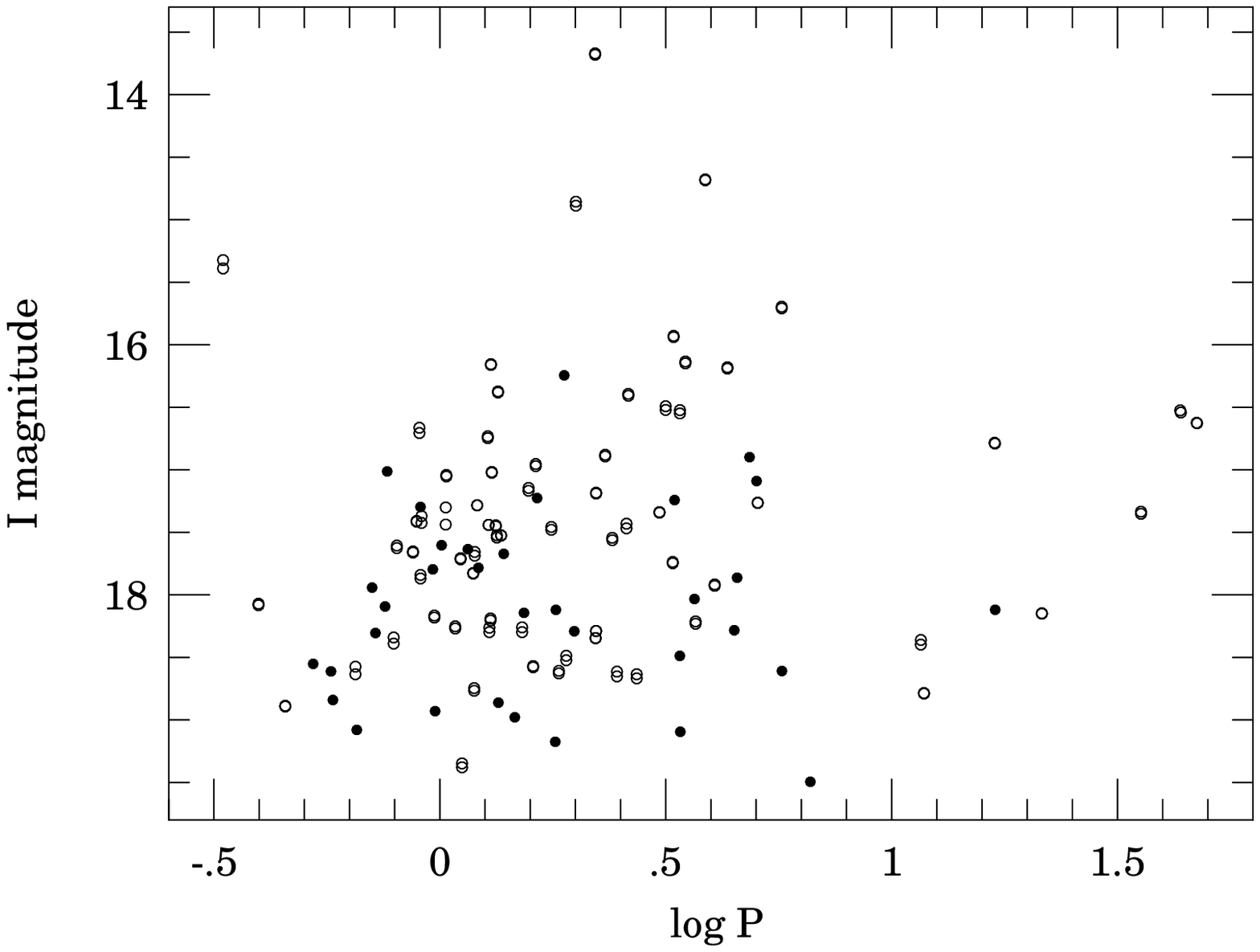,bbllx=30pt,bblly=40pt,bburx=520pt,bbury=440pt,width=10cm,clip=}}
\FigCap{Brightness \vs $\log P$ diagram for eclipsing stars which should
be paired up with stars located in neighboring fields. Open circles
mark identified stars while filled circles those which remained unpaired. }
\end{figure}
\begin{figure}[htb]
%\vskip5mm
\centerline{\psfig{figure=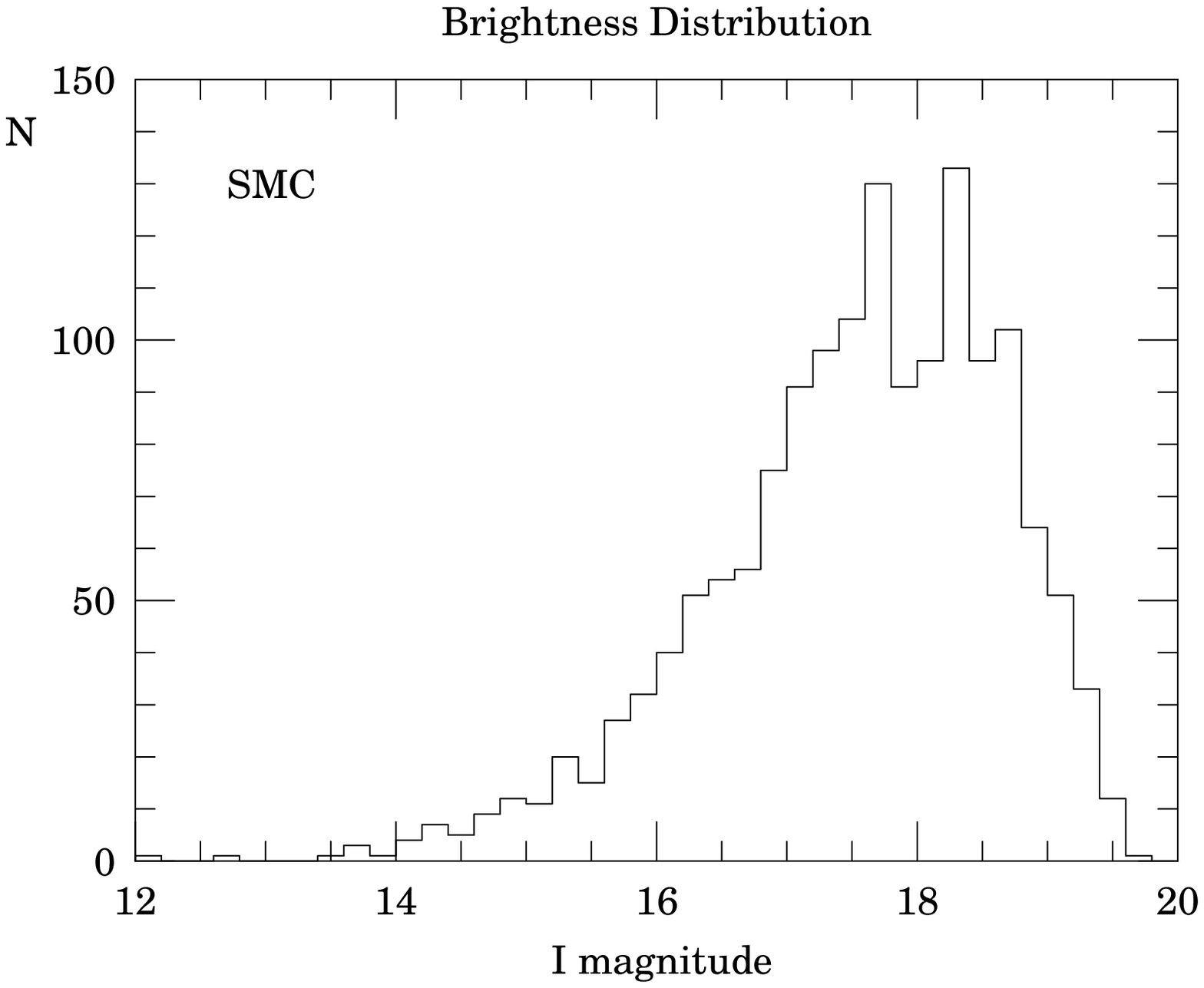,bbllx=20pt,bblly=40pt,bburx=520pt,bbury=410pt,width=10cm,clip=}}
\FigCap{Histogram of brightness of eclipsing stars in
0.2~mag bins.}
\end{figure} 
Fig.~4  presents the diagram of brightness of the system \vs $\log P$
for stars used in our tests. Open circles mark objects which were
identified in neighboring fields. Filled circles show missed pairs. It
is clear that completeness is a function of magnitude and most unpaired
stars are objects fainter than ${I\approx18}$~mag. Fig.~5 presents the
histogram of brightness of all eclipsing binaries. Figs.~4 and 5
indicate that  completeness is likely to be better than 90\% for objects
brighter than  ${I<17.5}$~mag and it falls gradually to zero for stars
fainter than ${I\approx19.5}$~mag. 

One of the most direct and important applications of the catalog is
selection of systems suitable for distance determination. The catalog
contains many well detached eclipsing systems which, when precise
spectroscopy is available, should provide accurate distances. The best
systems for distance determination should possess deep and narrow
eclipses assuring that both components are spherical and well separated.
To facilitate selection we extracted from the catalog 153 object sample
of detached eclipsing systems with well defined eclipses of similar
depth (depth of the secondary -- shallower -- eclipse larger than half
of the primary one), possibly flat light curve between eclipses and good
photometry. In Fig.~6 we plot depth of the secondary eclipse as a
function of full width at half depth of the secondary eclipse measured
as fraction of the orbital period. It can be seen from Fig.~6 that the
typical width of the secondary eclipse is about $0.06-0.07 P$ with depth
 0.3--0.5~mag. The deepest secondary eclipse systems include SMC$\_$SC5
283079, SMC$\_$SC2 11454 and  SMC$\_$SC6 251588 with depth about
$0.65-0.7$~mag. One should remember that depth for two identical star
system with total eclipses would be 0.75~mag. The most narrow eclipses
are those of SMC$\_$SC6 242498, SMC$\_$SC5 180516, SMC$\_$SC4 192903
($P=103$~days), SMC$\_$SC6 11143 and SMC$\_$SC4 14924 -- all below $0.03
P$. The list of all selected detached systems is available from the OGLE
Internet archive.

\begin{figure}[htb]
\vskip-5mm
\centerline{\psfig{figure=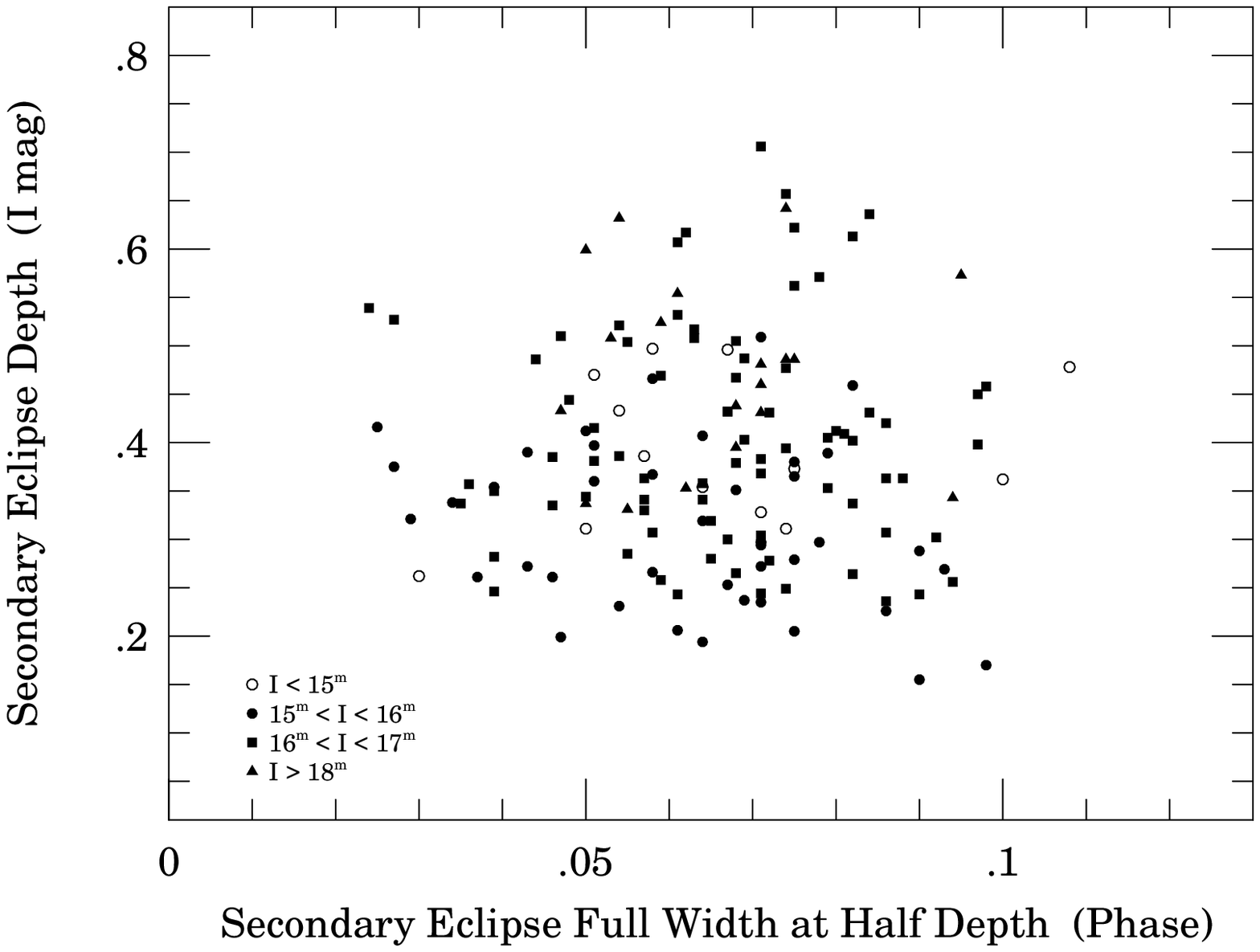,bbllx=30pt,bblly=40pt,bburx=520pt,bbury=440pt,width=10cm,clip=}}
\FigCap{Depth \vs full width at half depth of the secondary eclipse of well 
detached eclipsing systems.}
\end{figure}

Summarizing, we provide a reasonably complete sample of eclipsing binary
stars  from the SMC. It can be used for statistical analysis of the SMC
stars and  many other projects requiring multiband photometry of
eclipsing stars. Similar  catalog of eclipsing systems in the LMC will
be released by the OGLE-II  collaboration in the near future. 

The catalog, including {\it BVI}-band photometry data and finding charts
is  available to the astronomical community from the OGLE
archive:\newline  {\it http://www.astrouw.edu.pl/\~{}ftp/ogle} or its
mirror \newline {\it http://www.astro.princeton.edu/\~{}ogle}

\Acknow{We would like to thank Prof.\ Bohdan Paczy\'nski for many
discussions  and help at all stages of the OGLE project. The paper was
partly supported by  the Polish KBN grant 2P03D00814 to A.\ Udalski. 
Partial support for the OGLE  project was provided with the NSF grant
AST-9530478 to B.~Paczy\'nski. We  acknowledge usage of The Digitized
Sky Survey which was produced at the Space  Telescope Science Institute
based on photographic data obtained using The UK  Schmidt Telescope,
operated by the Royal Observatory Edinburgh.}

\newpage

\renewcommand{\TableFont}{\scriptsize}
%\vspace*{-25pt}
\setcounter{table}{12}
\MakeTable{
c@{\hspace{3pt}}
r@{\hspace{4pt}}
c@{\hspace{4pt}}
c@{\hspace{3pt}}
r@{\hspace{30pt}}
c@{\hspace{3pt}}
r@{\hspace{4pt}}
c@{\hspace{4pt}}
c@{\hspace{3pt}}
r}{12.5cm}{Cross-identification of stars detected in overlapping regions}
{
SMC$\_$SC1 &  97290 & $\leftrightarrow$ &  SMC$\_$SC2 &  2742 & SMC$\_$SC5 & 300848 & $\leftrightarrow$ &  SMC$\_$SC6 & 56181 \\  
SMC$\_$SC1 & 108856 & $\leftrightarrow$ &  SMC$\_$SC2 & 10301 & SMC$\_$SC5 & 300727 & $\leftrightarrow$ &  SMC$\_$SC6 & 61509 \\  
SMC$\_$SC2 &  86546 & $\leftrightarrow$ &  SMC$\_$SC3 &  8986 & SMC$\_$SC5 & 311566 & $\leftrightarrow$ &  SMC$\_$SC6 & 67289 \\  
SMC$\_$SC2 &  88691 & $\leftrightarrow$ &  SMC$\_$SC3 & 13032 & SMC$\_$SC5 & 311654 & $\leftrightarrow$ &  SMC$\_$SC6 & 67392 \\  
SMC$\_$SC3 & 193416 & $\leftrightarrow$ &  SMC$\_$SC4 &  8326 & SMC$\_$SC6 & 263814 & $\leftrightarrow$ &  SMC$\_$SC7 &   540 \\  
SMC$\_$SC3 & 194000 & $\leftrightarrow$ &  SMC$\_$SC4 &  8987 & SMC$\_$SC6 & 268972 & $\leftrightarrow$ &  SMC$\_$SC7 &  9072 \\  
SMC$\_$SC3 & 198240 & $\leftrightarrow$ &  SMC$\_$SC4 & 11552 & SMC$\_$SC6 & 272453 & $\leftrightarrow$ &  SMC$\_$SC7 & 13487 \\  
SMC$\_$SC3 & 202981 & $\leftrightarrow$ &  SMC$\_$SC4 & 15096 & SMC$\_$SC6 & 277060 & $\leftrightarrow$ &  SMC$\_$SC7 & 18332 \\  
SMC$\_$SC3 & 208373 & $\leftrightarrow$ &  SMC$\_$SC4 & 22592 & SMC$\_$SC6 & 292086 & $\leftrightarrow$ &  SMC$\_$SC7 & 32419 \\  
SMC$\_$SC3 & 213289 & $\leftrightarrow$ &  SMC$\_$SC4 & 26278 & SMC$\_$SC6 & 311434 & $\leftrightarrow$ &  SMC$\_$SC7 & 47309 \\  
SMC$\_$SC3 & 213548 & $\leftrightarrow$ &  SMC$\_$SC4 & 26565 & SMC$\_$SC6 & 307537 & $\leftrightarrow$ &  SMC$\_$SC7 & 47996 \\  
SMC$\_$SC3 & 217692 & $\leftrightarrow$ &  SMC$\_$SC4 & 29281 & SMC$\_$SC6 & 319956 & $\leftrightarrow$ &  SMC$\_$SC7 & 61679 \\  
SMC$\_$SC3 & 218379 & $\leftrightarrow$ &  SMC$\_$SC4 & 29615 & SMC$\_$SC7 & 233103 & $\leftrightarrow$ &  SMC$\_$SC8 &  7991 \\  
SMC$\_$SC3 & 222224 & $\leftrightarrow$ &  SMC$\_$SC4 & 29960 & SMC$\_$SC7 & 236364 & $\leftrightarrow$ &  SMC$\_$SC8 & 11553 \\  
SMC$\_$SC4 & 147702 & $\leftrightarrow$ &  SMC$\_$SC5 &  3407 & SMC$\_$SC7 & 237128 & $\leftrightarrow$ &  SMC$\_$SC8 & 15581 \\  
SMC$\_$SC4 & 154601 & $\leftrightarrow$ &  SMC$\_$SC5 & 12201 & SMC$\_$SC7 & 247769 & $\leftrightarrow$ &  SMC$\_$SC8 & 22435 \\  
SMC$\_$SC4 & 160383 & $\leftrightarrow$ &  SMC$\_$SC5 & 17284 & SMC$\_$SC7 & 255621 & $\leftrightarrow$ &  SMC$\_$SC8 & 30634 \\  
SMC$\_$SC4 & 163681 & $\leftrightarrow$ &  SMC$\_$SC5 & 21245 & SMC$\_$SC7 & 255665 & $\leftrightarrow$ &  SMC$\_$SC8 & 30691 \\  
SMC$\_$SC4 & 167321 & $\leftrightarrow$ &  SMC$\_$SC5 & 26819 & SMC$\_$SC7 & 259593 & $\leftrightarrow$ &  SMC$\_$SC8 & 35214 \\  
SMC$\_$SC4 & 171629 & $\leftrightarrow$ &  SMC$\_$SC5 & 38169 & SMC$\_$SC7 & 263170 & $\leftrightarrow$ &  SMC$\_$SC8 & 38913 \\  
SMC$\_$SC4 & 175149 & $\leftrightarrow$ &  SMC$\_$SC5 & 38079 & SMC$\_$SC8 & 170230 & $\leftrightarrow$ &  SMC$\_$SC9 &    99 \\  
SMC$\_$SC4 & 175576 & $\leftrightarrow$ &  SMC$\_$SC5 & 39026 & SMC$\_$SC8 & 174197 & $\leftrightarrow$ &  SMC$\_$SC9 &  4030 \\  
SMC$\_$SC4 & 176156 & $\leftrightarrow$ &  SMC$\_$SC5 & 44249 & SMC$\_$SC8 & 174291 & $\leftrightarrow$ &  SMC$\_$SC9 &  4108 \\  
SMC$\_$SC4 & 187310 & $\leftrightarrow$ &  SMC$\_$SC5 & 61542 & SMC$\_$SC8 & 198652 & $\leftrightarrow$ &  SMC$\_$SC9 & 30310 \\  
SMC$\_$SC4 & 198836 & $\leftrightarrow$ &  SMC$\_$SC5 & 74923 & SMC$\_$SC8 & 198995 & $\leftrightarrow$ &  SMC$\_$SC9 & 30455 \\  
SMC$\_$SC5 & 255984 & $\leftrightarrow$ &  SMC$\_$SC6 &  5337 & SMC$\_$SC8 & 207567 & $\leftrightarrow$ &  SMC$\_$SC9 & 39127 \\  
SMC$\_$SC5 & 257236 & $\leftrightarrow$ &  SMC$\_$SC6 &  6812 & SMC$\_$SC8 & 207699 & $\leftrightarrow$ &  SMC$\_$SC9 & 39292 \\  
SMC$\_$SC5 & 261267 & $\leftrightarrow$ &  SMC$\_$SC6 & 11806 & SMC$\_$SC9 & 144175 & $\leftrightarrow$ &  SMC$\_$SC10 &  114 \\ 
SMC$\_$SC5 & 265970 & $\leftrightarrow$ &  SMC$\_$SC6 & 17345 & SMC$\_$SC9 & 147421 & $\leftrightarrow$ &  SMC$\_$SC10 & 6388 \\ 
SMC$\_$SC5 & 266131 & $\leftrightarrow$ &  SMC$\_$SC6 & 22883 & SMC$\_$SC9 & 152769 & $\leftrightarrow$ &  SMC$\_$SC10 &11958 \\ 
SMC$\_$SC5 & 271839 & $\leftrightarrow$ &  SMC$\_$SC6 & 29314 & SMC$\_$SC9 & 161377 & $\leftrightarrow$ &  SMC$\_$SC10 &19778 \\ 
SMC$\_$SC5 & 277666 & $\leftrightarrow$ &  SMC$\_$SC6 & 29828 & SMC$\_$SC9 & 175323 & $\leftrightarrow$ &  SMC$\_$SC10 &33878 \\ 
SMC$\_$SC5 & 283889 & $\leftrightarrow$ &  SMC$\_$SC6 & 36041 & SMC$\_$SC9 & 175488 & $\leftrightarrow$ &  SMC$\_$SC10 &33982 \\ 
SMC$\_$SC5 & 300752 & $\leftrightarrow$ &  SMC$\_$SC6 & 55836 & SMC$\_$SC10& 118942 & $\leftrightarrow$ &  SMC$\_$SC11 &13386 \\ 
}


\begin{references}
\refitem{Alcock, C. \etal}{1997}{\AJ}{114}{326}
\refitem{Grison, P. \etal}{1994}{\AAS}{109}{447}
\refitem{Guinan, E.F. \etal}{1998}{\ApJL}{509}{L21}
\refitem{Kaluzny, J., Stanek, K.Z., Krockenberger, M., Sasselov, D.D.,
Tonry, J.L., and Mateo, M.}{1998}{\AJ}{115}{1016}
\refitem{Paczy\'nski, B.}{1997}{~}{~}{in: "The Extragalactic Distance Scale 
STScI Symposium", Baltimore, Cambridge University Press, 273 
(astro-ph/9608094)}
\refitem{Schwarzenberg-Czerny, A.}{1989}{\MNRAS}{241}{153}
\refitem{Stanek, K.Z., Kaluzny, J., Krockenberger, M., Sasselov, D.D.,
Tonry, J.L., and Mateo, M.}{1998}{\AJ}{115}{1894}
\refitem{Stebbins, J.}{1910}{\ApJ}{32}{185}
\refitem{Udalski, A., Olech, A., Szyma\'nski, M.,  Ka{\l}u{\.z}ny, J., 
Kubiak, M., Mateo, M., Krzemi{\'n}ski, W., and Stanek, K.Z.}
{1997a}{\Acta}{47}{1} 
\refitem{Udalski, A., Kubiak, M., and Szyma\'nski, M.}{1997b}{\Acta}{47}{319}
\refitem{Udalski, A., Szyma\'nski, M., Kubiak, M., Pietrzy\'nski, G.,
Wo\'zniak, P., and {\.Z}ebru\'n, K.}{1998a}{\Acta}{48}{147}
\refitem{Udalski, A., Pietrzy\'nski, G., Wo\'zniak, P., Szyma\'nski, M.,
Kubiak, M.,  and {\.Z}ebru\'n, K.}{1998b}{\ApJL}{509}{L25}
\end{references}
\end{document}